\documentstyle[11pt,blois,epsfig]{article}

\def\gtwid{\mathrel{\raise.3ex\hbox{$>$\kern-.75em\lower1ex\hbox{$\sim$}}}}
\def\ltwid{\mathrel{\raise.3ex\hbox{$<$\kern-.75em\lower1ex\hbox{$\sim$}}}}
\def\\{\hfil\break}
\def\ie{{\it i.e.\ }}

\def\etal{{\it et al.\ }}

\newcommand{\vs}{v_{12}}
\newcommand{\se}{\sigma_8}
\newcommand{\Mpc}{\, h^{-1} \, {\rm Mpc}} 
\def\prl{\em Phys. Rev. Lett.}

\def\apj{{\em ApJ} }
\def\apjl{{\em ApJ Lett.} }
\def\apjs{{\em ApJ Supl.} }
\def\aj{{\em AJ} }
\def\mnras{{\em MNRAS} }

\def\be{\begin{equation}}
\def\ee{\end{equation}}
\def\bea{\begin{eqnarray}}
\def\eea{\end{eqnarray}}

\begin{document}
\vspace*{4cm}
\title{Velocity Fields as a Probe of Cosmology}

\author{ Hume A. Feldman}
\address{$^\dagger$Department of Physics \& Astronomy,
University of Kansas, Lawrence, KS 66045, USA}

\maketitle\abstracts{
Analyses of peculiar velocity surveys face several challenges, including
low signal--to--noise in individual velocity measurements and the
presence of small--scale, nonlinear flows.  I will present three new analyses 
that attempt to address these inherent problems. The first is geared towards the better understanding of the estimated errors in the surveys, specifically sampling errors, and the resolution of the seeming disagreements between the surveys. Another develops a new statistic that does not suffer from the usual problems and gives robust results that are galaxy--morphology and distance--estimator independent. The third introduces a formalism that allows for the accounting of most of the non--linear signal whereby the signal to noise is increased and small--scale aliasing is removed.
}

\section{Introduction}
\label{sec-intro}

The best developed class of theories of structure formation is based on
the assumption of small amplitude Gaussian density perturbations which
grew and condensed out, becoming, by gravitational instability,
galaxies, clusters of galaxies, superclusters or voids.  The study of
the evolution of density perturbations is the basis of all theories
based on gravitational instability~\cite{LSS,sz89}.  Recently, this class of theories was given a dramatic
boost by the detection of appropriate temperature fluctuations in the
cosmic microwave background. Further, recent work on the
bispectrum measurement in the IRAS redshift catalogs~\cite{bis2} shows strong signature of gravitational instability.

Statistical studies of the peculiar velocity field (those velocities in excess of the pure Hubble
flow) based on the predictions of the gravitational instability scenario
can potentially address many fundamental cosmological questions.
Although in principle the galactic peculiar (local) velocity field holds
a great promise as a direct probe of the underlying mass distribution,
in practice the extraction of the information is difficult and fraught
with both observational and theoretical pitfalls. Velocity surveys have
irregular geometries and their boundary conditions are not usually
well--known. Further, they are somewhat shallow and sample the volume
discretely, non--uniformly and sparsely. Theoretically, the mapping from
velocity to density is complicated by non--linear effects which
necessitates various approximation schemes~\cite{Z70}. Indeed, small--scale aliasing and incomplete cancellations~\cite{wf95,fw98} introduce spurious
noise which masquerades as large scale signal. These effects are
difficult to disentangle and thus the resulting information is
unreliable~\cite{fw94,slp95}. The bottom line with respect to peculiar velocity surveys
is that it is very difficult to extract cosmological
information since the large--scale signal has a significant small--scale
noise which is virtually impossible to model accurately. 

Estimating cosmological parameters may be done by studying the
large scale motions of galaxies.  One advantage of this method is that
the large scale velocity field probes the matter distribution in the
Universe directly, and not merely the light distribution as redshift
surveys do.  However, to measure the velocity field
one needs to make accurate distance measurements, which has proven to be
quite difficult~\cite{jacoby}.  The errors in distance estimates are typically some
fraction of the redshift of the sample points, which in the case of
distant objects can mean that the errors are larger than the peculiar
velocity being measured.  This is partially rectified by measuring only
the lowest moment of the velocity field, namely the bulk flow.  Since
the bulk flow is in a sense an average of the velocities in the sample,
its error is reduced over that of an individual measurement by the
square root of the number of objects~\cite{LP,RPK}.  The idea here is that in calculating low--order
moments the small scale modes will be averaged out, so that the values
of these moments will reflect only large--scale motion.  It has been
shown, however, that the sparseness of peculiar velocity data can lead
to small--scale modes making a significant contribution to low--order
moments through incomplete cancellation ~\cite{fw94,fw98}.  Another drawback of this
approach is that it utilizes only a fraction of the available
information.

An alternative method is to perform a likelihood analysis using all of
the velocity information~\cite{jk95}.  An obvious danger here is that
retaining small--scale, nonlinear contributions to the velocities can
lead to unpredictable biases which can skew the results~\cite{C&E}.  This method also has the disadvantage of becoming unwieldy for surveys larger than about a thousand objects.  While
advances in computing will make this less of a problem in the future,
clearly a less time--intensive method is desirable.

The non--linear aliasing and incomplete cancellations inherent
in the surveys effectively defied all attempts to extract robust cosmological
information from the data. The remedies proposed in the literature, POTENT~\cite{potent}; denser directional surveys~\cite{fw98} and others where not satisfactory. Smoothing over the
problems by introducing some averaging schemes did not work consistently. Smoothing,
including Weiner Filtering~\cite{wiener} provide formalisms
that may or may not remove small scale irregularities. The problem is
that since smoothing operations provides little or no control over what
we remove and what we keep, that it is impossible to state with any
degree of certainty whether the smoothed field retains the large scale
signal while removing the potentially significant small scale noise.

Here I would like to discuss three programs that have been developed to address different aspects of  the problems discussed above. The first attempts to better understand the error estimations inherent in velocity field surveys and to develop a scheme to compare them and see if the surveys themselves are compatible with each other within their errors and with the power spectrum of density fluctuations derived by other techniques. Another searches for a statistic that is robust and does not depend on the particular distance indicator or morphology. The last tries to deal with the problems head on, that is, to look for a statistical way to rid the surveys of their non--linearities in such a way that the large--scale signal remains intact whereas the small--scale "noise" is being removed.

One idea proposed is to develop a more realistic error estimation for existing surveys, in particular, the understanding of the sampling errors~\cite{hudson03} allows for consistent bulk flow estimates for most of the bulk flow measurements within the errors (see Table~\ref{tab:bulk} below). Additionally, we get an idea of how much correlation we expect between the different catalogs for a given power spectrum by calculating the normalized expectation value for their dot--product which should be close 
to $1$ for highly correlated surveys, zero for those that are completely uncorrelated, and $-1$ if there is a high degree of anti-correlation~\cite{wf95,hudson99,CF}.   As we see elsewhere in this volume~\cite{hudson03} the bulk flow measurements of various independent surveys are fairly correlated and a reasonably consistent bulk flow vector emerges from them that is $380\pm80$km/s  towards $l=290^o$ and $b=10^o$.

\begin{table}[h]
\centerline{Recent Large Scale Bulk Flow Measurements}
\vspace{0.4cm}
\begin{center}
\begin{tabular}{|l|l|r|r|r|r|r|}
\hline     
Survey                           & Method & N       & v$_{\rm pec}$ & Total  & l      & b \\
                                        &                &          & km/s                  & error  &        &    \\
\hline
\hline
LP~\cite{LP}                   & BCG     & 119  & 830                & 370      & 330 & 39 \\ \hline
Willick~\cite{shellflow}  & TF         & 15    & 1060             & 670      & 275 & 28 \\ \hline
SC~\cite{SC}                   & TF         & 63    & 120               & 295       & 10  & 310 \\ \hline
SMAC~\cite{hudson99}& FP        & 56    & 690                & 380      & 260 & -1 \\ \hline
EFAR~\cite{EFAR}        & FP         & 49    & 630                & 410     & 50 & 10 \\ \hline
SN~\cite{Tonry}              & SNIa     & 65   & 610                & 330      & 313 & 9 \\ \hline
\end{tabular}
\caption{When we take into account both the total error, \ie random {\it and} sampling errors, most bulk flow measurements agree with each other. For a more detailed analysis see Hudson, M., 2003, In this volume.}
\label{tab:bulk}
\end{center}
\end{table}

Another program that seeks to address the inconsistencies between various velocity surveys results  develops a statistic that is not as susceptible to small--scale aliasing and incomplete cancellations. In series of recent papers we introduced a new dynamical estimator of the $\Omega_m$ parameter, the dimensionless density of the
nonrelativistic matter in the universe.  We use the so called {\it streaming velocity}, or the mean relative peculiar velocity for galaxy pairs, $\vs(r)$, where $r$ is the pair separation~\cite{LSS}. It is measured directly from peculiar velocity surveys, without the noise-generating spatial differentiation, used
in reconstruction schemes, like POTENT (see Courteau \etal 2000 and references therein).
In the first paper of the series~\cite{rj98}, we derived an equation,
relating $\vs(r)$ to $\Omega_m$ and  the two-point correlation function of mass
density fluctuations, $\xi(r)$.  Then, we showed that $\vs$ and $\Omega_m$
can be estimated from mock velocity surveys~\cite{pairwise1}, from real data: the
Mark III survey~\cite{pairwise2} and finally~\cite{pairwise3} a comparison of various independent surveys that use different standard candles, galaxy morphologies and surveying techniques. Whenever a new statistic is introduced, it is of particular importance that it passes the test of reproducibility. Our results pass these tests: the $\vs(r)$ measurements are galaxy morphology-- and distance indicator--independent~\cite{rj03}.
 
Using mean relative peculiar velocity measurements for pairs of galaxies, we estimate the cosmological density parameter $\Omega_m$ and the amplitude of density fluctuations $\se$.   Our results suggest that our statistic is a robust and reproducible measure of  the mean pairwise velocity and
thereby the $\Omega_m$ parameter.  We get $\Omega_m = 0.30^{+0.17}_{-0.07}$ and $\se = 1.13^{+0.22}_{-0.23}$.  These estimates do not depend on prior assumptions on the adiabaticity of the initial density fluctuations, the ionization history, or the values of other cosmological parameters.

\begin{figure}[h]
\begin{center}
\psfig{figure=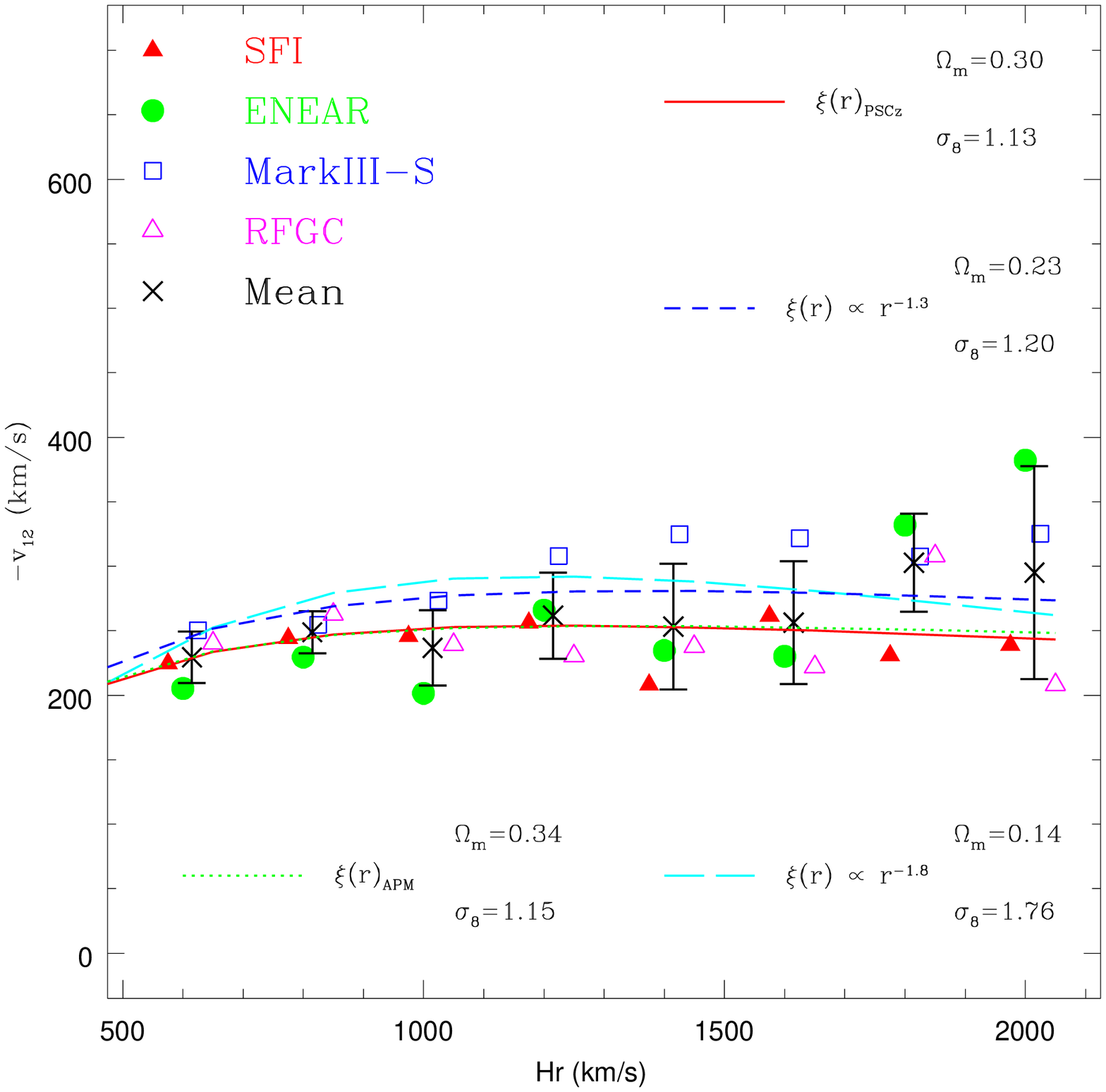,height=2.3in}\qquad\qquad\qquad
\psfig{figure=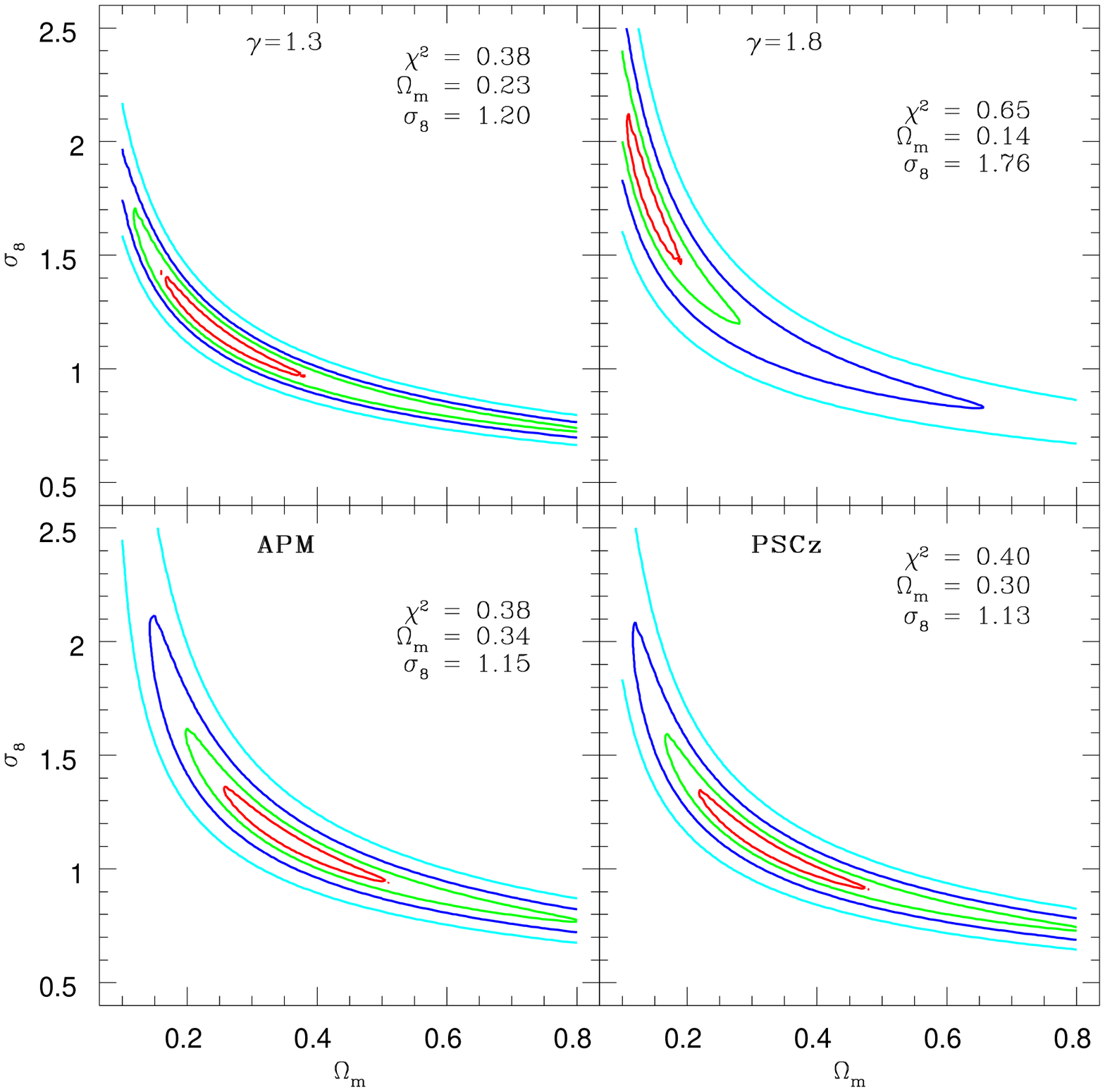,height=2.3in}
\end{center}
\caption{
(left panel) The crosses and the associated error bars show 
the weighted mean pairwise velocity, obtained by averaging over four
surveys. Individual survey data points are also shown; we have
suppressed their error bars for clarity. These direct measurements
of $ \vs$ are compared to four $ \vs(r)$ curves, derived by assuming 
four different models of  $ \xi(r)$. The labels 
identify best fit $ \Omega_m$ and $ \se$ parameters.\hfill\break
Figure 2: (right panel) The results of the maximum likelihood analysis. The 
upper panels show results for power-law toy models, while the
bottom panels are based on realistic representations of 
observations: the APM and PSCz data, respectively. 
Likelihood peak coordinates and the values of $ \chi^2$ for 
each model are also indicated. The innermost contours define 
the 68\%, or 1-$\sigma$ areas around the peaks.  The remaining 
nested contours show the 2, 3 and 4-$\sigma$ boundaries.\hfill\break
}
\label{fig1}
\end{figure}

In figure 1~\cite{pairwise3} we see that the claim that the $\vs$ statistic is independent of the galaxy morphology or the standard candle is correct. We show here four independent surveys, The Mark III~\cite{m3}, a TF compilations of 2437 spiral galaxies.  The SFI~\cite{SFI} catalogue contains 1300 late type spiral galaxies with $I$-band TF distance estimates. The ENEAR~\cite{ENEAR} survey with 1359 early type with $D_n$-$\sigma$ measured distances. The RFGC~\cite{RFGC} catalogue provides a list of radial velocities, HI line widths, TF distances and peculiar velocities of 1327 spiral galaxies that was compiled from observations of flat galaxies performed with the 305\,m telescope at Arecibo.  As can clearly be seen, all measurements agree with each other and thus allows us to combine the results and obtain a mean $\vs(r)$ with smaller errors than for each individual survey. Further, we see that since all surveys give us identical results within the errors, ther is no sign for velocity bias.

In figure 2 we see the results of our estimates for the parameters, $\Omega_m$ and $\se$. When we use a correlation function obtained observationally, such as from the IRAS--PSCz survey~\cite{HT} or the one from the APM~\cite{gj01} we get results as those quoted above. However, if we use a correlation function $\xi(r)\propto r^{-1.8}$ (a seeming favourite in some quarters, through it is not compatible with either the PSCz or the APM surveys) we get very small $\Omega_m$ and very large $\se$. Thus this statistic can also act as a regularizing diagnostic for the form the correlation function takes around the scale it is sensitive (on the order of $10 \Mpc$).

The last program I present is one that is designed to remove the non--linear signal from the survey. The method utilizes Karhunen--Lo\`eve
methods of data compression to construct a set of moments out of the
velocities which are minimally sensitive to small scale power; these
moments can then be used in a likelihood analysis.  Overall sensitivity
of the set of moments to small scales is quantified, and can be
controlled through the number of moments retained in the analysis.
Since the number of moments kept is typically much smaller than the
number of velocities in the survey, this method has the added advantage
of being much more efficient than a full analysis of the data.

Karhunen--Lo\`eve methods~\cite{KK,KS} have recently become popular in
cosmology. A general discussion of their use in the analysis of large
data sets was done here~\cite{TTH}.  In addition, these
methods have been applied to the Las Campanas Redshift Survey~\cite{matsubara00}, to velocity field surveys~\cite{hz00}, and to the decorrelation of the power
spectrum~\cite{H00,HT00}. Although we use the same general method, our
take on the formalism is quite different.  Taking advantage of the
compression techniques and the Fisher information matrix,
we filter out small--scale, nonlinear velocity modes and retain only
information regarding the large--scale modes.

Technical details of the formalism are given elsewhere~\cite{PI,PII,feldman03}. Here I would only like to present some of the results. The formalism allows the diagonalization of the covariance matrix with eigenvalues whose amplitudes are proportional to the sensitivity to small--scale modes. A set of optimal moments constructed as linear combinations of velocities which are minimally sensitive to
small scales.  The overall sensitivity of a set of moments to small scales can be quantified and controlled through the choice of the number of moments retained.

In Fig.~3 we show the window functions
for selected moments in order of
increasing eigenvalue. This demonstrates that selecting moments that are least sensitive to
small scales generally results in moments that are most
sensitive to large scales; window functions of moments with larger
eigenvalues have successively larger amplitudes on nonlinear scales as expected.
Thus the information contained in large eigenvalue moments comes mostly
from scales where fluctuations are nonlinear and should not be included
in a linear analysis.

\setcounter{figure}{2}
\begin{figure}[t]
\begin{center}
\psfig{figure=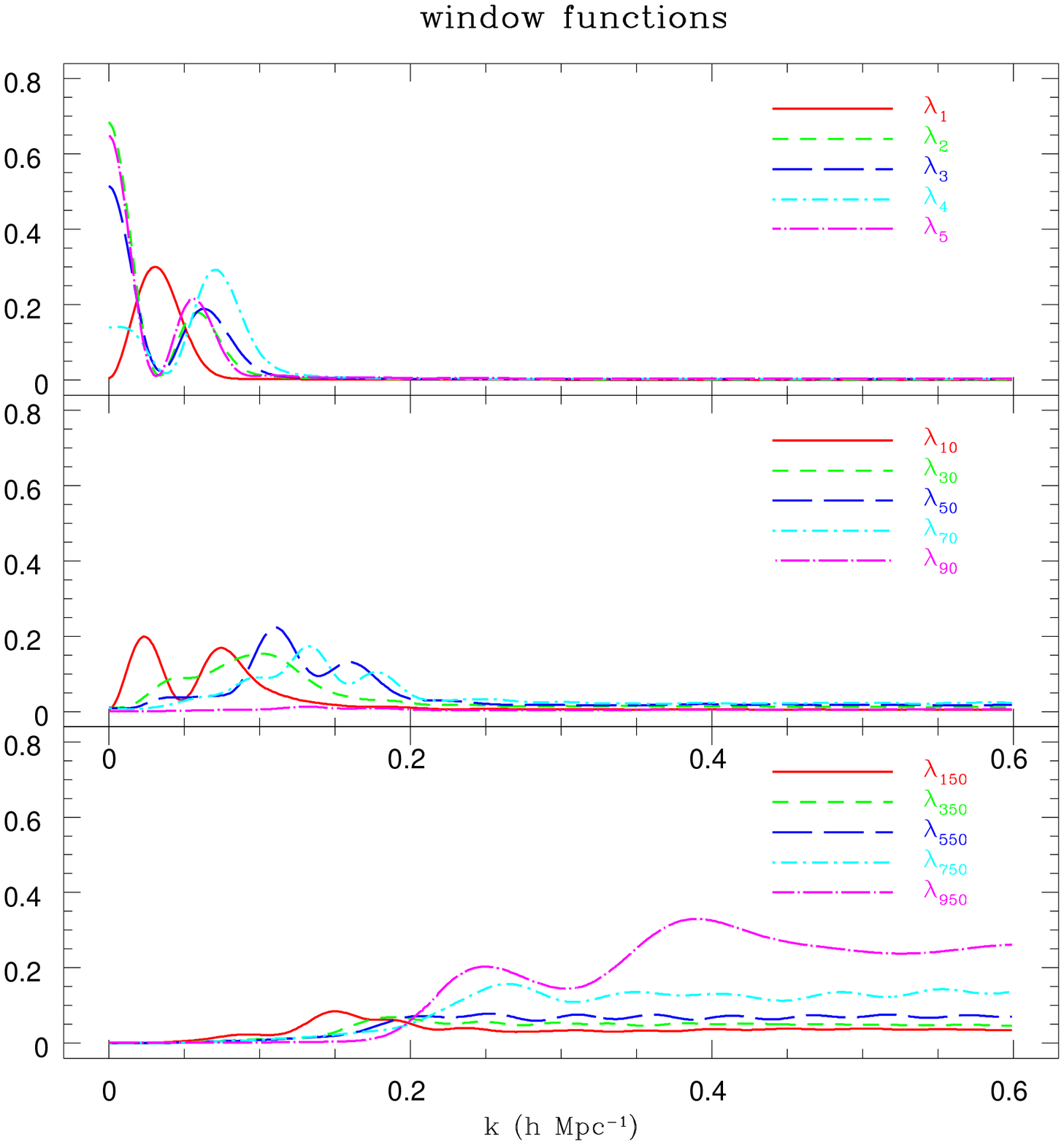,height=2.3in}\qquad\qquad\qquad
\psfig{figure=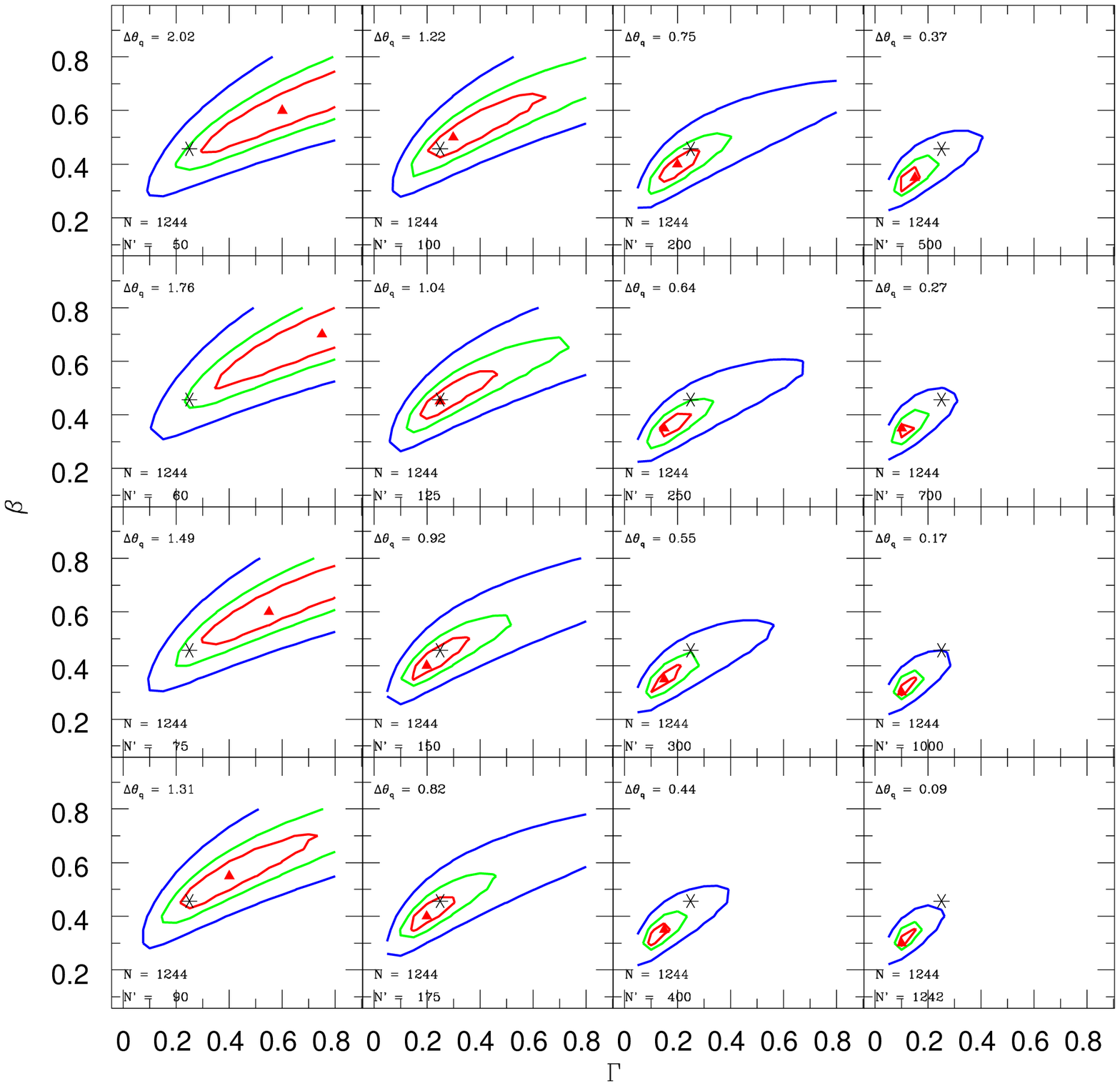,height=2.3in}
\end{center}
\caption{
(left panel) The window functions in arbitrary units. The
top panel shows the window functions associated with the five smallest
eigenvalues. The center panel shows the window functions associated with
five somewhat larger eigenvalues, while the bottom panel shows the
window functions selected from the entire range of large eigenvalues. For each
panel we show the eigenvalue rank for each window function. \hfill\break
Figure 4: (right panel) Likelihood vs. $\Gamma$ and $\beta$ for the
simulated catalog for different
$N^{\prime}$, the number of moments kept; we also show the value of
$\Delta\theta_q$.  The panels show the maximum likelihood value
(solid triangles) and the contours corresponding to $0.5,0.1$ and $0.01$
of the maximum values. The asterisk in each panel is the ``true'' value
of $\Gamma=0.25$ and $\beta=0.457$ for the simulation. Increasing the
number of modes $N^\prime$ improves the accuracy of the maximum
likelihood values up to a point, but the inclusion of large eigenvalue
modes that carry information mostly from nonlinear scales can skew the
result away from the true value. \hfill\break}
\label{fig2}
\end{figure}

In Fig.~4 we show the results of the likelihood
analysis on a typical catalog for various $N^{\prime}$.  For reference, we also give the value of
$\Delta\theta_q$ for each $N^{\prime}$.  We see that in this case, inclusion of all of the information
leads to the location of the maximum likelihood being skewed
away from the true values. However, when higher order moments are discarded, the location
of the maximum likelihood corresponds well with the true values.
The fact that 
the discarding of higher order moments leads to a much better
agreement between the maximum likelihood location and the true
values is a good indication that our analysis method is
effectively removing small--scale, nonlinear velocity
information.    

Although proper distance surveys present many challenges to cosmologists, they are quite well suited to the extraction of cosmological information provides they are handled with care and caution. In this conference I and my colleagues showed that with the right approaches we can extract interesting and important information and that techniques we have developed should be added to the cosmological toolbox. 

\section*{Acknowledgments}

I wish to acknowledge support
from the National Science Foundation under grant number AST--0070702,
the University of Kansas General Research Fund and the National Center
for Supercomputing Applications for allocation of computer time. This
research has been partially supported by the Lady Davis and Schonbrunn Foundation at
the Hebrew University, Jerusalem, Israel and by the Institute of
Theoretical Physics at the Technion, Haifa, Israel.


\section*{References}

\begin{thebibliography}{}

\bibitem{LSS} Peebles, P.J.E. 1980 {\it The Large Scale Structure of the
Universe} Princeton U. Press.

\bibitem{sz89} Shandarin, S. F., \& Zel'dovich, Ya. B. 1989,  Rev. Mod. Phys., 61, 185. 

\bibitem{bis2} Feldman, H.A. \& Frieman, J., Fry, J., Scoccimarro, R., 2001 \prl\  86, 1434.

\bibitem{Z70} Zel'dovich Ya.B. 1970, { A\&A}, 5, 84

\bibitem{wf95} Watkins, R. \& Feldman, H. A., 1995, \apjl\ 453 L72--76.

\bibitem{fw98} Feldman, H.A. \& Watkins, R. 1998, \apj\ 494 L129--132

\bibitem{fw94} Feldman, H. A. \& Watkins, R., 1994, \apjl\ 430 L17--20. 

\bibitem{slp95}Strauss, M., Cen, R., Ostriker, J. P., Postman, M. \& Lauer, T. 1995 \apj\ 444 507 

\bibitem{jacoby} Jacoby, G.H. \etal 1992, { Publ. Astron. Soc. Pac. 104}, 599.  

\bibitem{LP} Lauer, T. \& Postman, M. 1994, \apj 425 418.

\bibitem{RPK} Riess, A. G., Press, W. H., \& Kirshner, R. P. 1995, ApJ, 438, L17

\bibitem{jk95}Jaffe, A. \& Kaiser, N. 1995 \apj\ 255 26

\bibitem{C&E}Croft, R. \& Efstathiou, G., 1994, Potsdam Cosmology Workshop: astro--ph/9412024.

\bibitem{potent} Bertschinger, E. \& Dekel, A. 1989, \apjl\, 336, L5

\bibitem{wiener}Zaroubi, S., Hoffman, Y. \& Dekel, A.  \apj\ 520, 413

\bibitem{hudson03} Hudson, M., 2003, In this volume.

\bibitem{hudson99} Hudson, M., Smith, R., Lucey, J., Schlegel, D. \& Davies, R., 1999, \apjl\ L79.

\bibitem{CF} Courteau, Strauss, \& Willick, Eds., 2000, ASP
Conf. Ser. 201, Cosmic Flows

\bibitem{shellflow} S. Courteau , J.ÊWillick , M.ÊStrauss , D.ÊSchlegel , \& MÊPostman, 2000, \apj\ 544, 636

\bibitem{SC} Dale, D. A., Giovanelli, R. , Haynes, M. P., Campusano,
L. E., Hardy, E. and Borgani, S., 1999, \apjl, 510, L11.

\bibitem{EFAR} R. Saglia \etal, 1997, \apjs\ 109, 79.

\bibitem{Tonry} Tonry, J.~L. \etal, 2001, \apj, 594, 1.
 
\bibitem{SNIa}Riess, A. G., Press, W. H., \& Kirshner, R. P. 1996 473 88

\bibitem{rj98} Juszkiewicz, R., Springel, V. \& Durrer, R., 1999, \apj, 518, L25

\bibitem{pairwise1}  Ferreira, P. G., \etal, 1999, \apj, 515, L1

\bibitem{pairwise2}  Juszkiewicz, R., \etal, 2000, Sci, 287, 109

\bibitem{pairwise3} Feldman, H.A. \etal, 2003, \apjl\ 596 131L

\bibitem{rj03} Juszkiewicz, R., 2003, In this volume.

\bibitem{m3} Willick, J. A., \etal, 1997, \apjs, 109, 333

\bibitem{SFI} da Costa, L. N., \etal, 1996, \apj, 468, L5

\bibitem{ENEAR} da Costa, L. N., \etal, 2000, \aj, 120, 95

\bibitem{RFGC}  Karachentsev, I. D., \etal, 2000, Bull. Spec. Astrophys. Obs. N. Caucasus, 50, 5

\bibitem{HT} Hamilton, A. J. S., \& Tegmark, M. , 2002, \mnras, 330, 506

\bibitem{gj01}  Gazta{\~n}aga, E. \& Juszkiewicz, R., 2001, \apj, 558, L1

\bibitem{KK} Kenney, J.F., \& Keeping, E.S. 1954, {\it Mathematics of statistics}, Van Nostrand
company.

\bibitem{KS} Kendall, M. G. \& Stuart, A. 1969 {\it The advanced Theory of Statistics} Vol. 2, Grifin.

\bibitem{TTH} Tegmark, M., Taylor, A.N. \& Heavens, A.F., 1997, \apj\ 480 22

\bibitem{matsubara00} Matsubara, T., Szalay, A.S. \& Landy, S.D., 2000, \apj\ 535:L1

\bibitem{hz00} Hoffman, Y. \& Zaroubi, S., 2000, \apj\ 535 L5

\bibitem{H00} Hamilton, A., 2000 \mnras 312 257

\bibitem{HT00} Hamilton, A. \& Tegmark, M., 2000 \mnras 312 285

\bibitem{PI} Watkins, R., Feldman, H., Chambers, Gorman \& Melott, 2002,  \apj\ 564 534

\bibitem{PII}Feldman, H. A., Watkins, R., Melott, A. \& Chambers, W., 2003, astro--ph/0304316

\bibitem{feldman03} Feldman, Hume A., Watkins, R., Melott, A. \& Chambers, W., 2003, In this volume.

\end{thebibliography}
\end{document}